\documentclass[sn-mathphys-num]{sn-jnl}

\usepackage{graphicx}%
\usepackage{multirow}%
\usepackage{amsmath,amssymb,amsfonts}%
\usepackage{amsthm}%
\usepackage{mathrsfs}%
\usepackage[title]{appendix}%
\usepackage{xcolor}%
\usepackage{textcomp}%
\usepackage{manyfoot}%
\usepackage{booktabs}%
\usepackage{algorithm}%
\usepackage{algorithmicx}%
\usepackage{algpseudocode}%
\usepackage{listings}%

\theoremstyle{thmstyleone}%
\theoremstyle{thmstyletwo}%

\theoremstyle{thmstylethree}%

\begin{document}

\title[‌Spin Chain Integrability as a Supersymmetric Gauge Duality]{Spin Chain Integrability as a Supersymmetric Gauge Duality}


\author[1]{\fnm{Xiang-Mao} \sur{Ding}}\email{xmding@amss.ac.cn}

\author[2]{\fnm{Ting} \sur{Zhang}}\email{zhangting@pku.edu.cn}

\affil[1]{\orgdiv{State Key Laboratory of Mathematical Sciences (SKLMS), \\
Academy of Mathematics and Systems Science}, \\ 
\orgname{Chinese Academy of Sciences}, \orgaddress{\city{Beijing}, \postcode{100190}, \country{China}}}

\affil[2]{\orgdiv{School of Mathematical Sciences}, \orgname{Peking University}, \orgaddress{\city{Beijing}, \postcode{100871}, \country{China}}}


\abstract{We establish a novel correspondence between 4D $\mathcal{N}=1$ supersymmetric gauge theories on $D^2\times T^2$ and open $\text{XYZ}$ spin chains with generalized boundary conditions, extending beyond previous 3D Bethe/gauge duality frameworks. Our primary contribution is the rigorous construction of the first exact duality between 4D $\text{BCD}$-type gauge theories‌ and‌ the general-boundary $\text{XYZ}$ spin chains‌ governed by elliptic $R$-matrices. Crucially, this framework provides a universal mechanism for resolving supersymmetric gauge theory/spin-chain duality across dimensional hierarchies,
\begin{equation*}
\begin{aligned}
2D\ \mathcal N&=(2,2)\ \text{theory} && \leftrightarrow \text{XXX spin chain} \\
3D\ \mathcal N&=2\ \text{theory}      && \leftrightarrow \text{XXZ spin chain} \\
4D\ \mathcal N&=1\ \text{theory}      && \leftrightarrow \text{XYZ spin chain}
\end{aligned}
\end{equation*}
mediated through the $\Omega$-deformation parameter $\epsilon$.}




\maketitle

\tableofcontents

\section{Introduction}\label{s1}

The anisotropic $\text{XYZ}$ quantum spin chain is a prototypical $U(1)$ symmetry breaking quantum integrable system, plays significant roles in statistical mechanics, quantum magnetism, and mathematical physics \cite{Bax07, KBI93,SB13}. Baxter's foundational work \cite{Bax71a,Bax71b}, established its exact solution under periodic boundary conditions through inversion relations, with transfer matrix eigenvalues characterized by the $T\text{-}Q$ relation. Subsequent developments by 
Takhtadzhan and Faddeev \cite{TF79} via the quantum inverse scattering method (Algebraic Bethe Ansatz) \cite{KBI93,Tak99} systematized the solution space. 
Recent advances \cite{WYC15} further resolved the thermodynamic limit for 
non-diagonal boundary conditions with six free parameters.

The integrability structure of supersymmetric gauge theories, epitomized by the Bethe/Gauge correspondence, originated from Seiberg-Witten theory \cite{SW94a,SW94b}. This connection manifests through $R$-matrices encoding instanton dynamics in $\Omega$-backgrounds \cite{MO12,Nak13}, revealing profound links between gauge theory vacua and quantum many-body systems. A paradigmatic example is the equivalence between 2D $U(N)$ Yang-Mills theory and $N$-fermion dynamics on a one-dimensional ring.

In the literatures \cite{DZ23a,DZ23b,DZ23c,NS09a,NS09b,KZ21}, there is a one-to-one correspondence between the 2-, 3-, and 4-dimensional $U(N)$ supersymmetric gauge theory on $R^2\times T^2$ and the spin chain models with periodic boundary, the 2-, 3-dimensional type $\rm BCD$ gauge theory on $D^2\times S^1$ and the spin chain with diagonal boundary. And recently, the 4-dimensional type $\rm BD$ gauge theory on $D^2\times T^2$ was found to match the open $\text{XYZ}$ spin chain \cite{WZ23}. The Bethe gauge correspondence between type $\rm D$ relativistic Toda lattice and 5-dimensional $\mathcal{N} = 1$ folded theory was studied in \cite{LL24}.

In this paper, based on previous studies \cite{DZ23a,DZ23b,DZ23c}, we establish the correspondence between the 4D $\text{BCD}$-type supersymmetric gauge theory and $\text{XYZ}$ spin chain with general boundary conditions. On the integrable system side, we adopt the Bethe Ansatz equations (BAEs) derived in \cite{XCYW24}, expressed as
\begin{equation}\label{1}
    \begin{aligned}
      &-\frac{\sigma(i(u_j-\tfrac{\eta i}{2})+\tfrac{1}{2})\sigma(i(u_j-\tfrac{\eta i}{2})+\tfrac{\tau}{2})\sigma(i(u_j-\tfrac{\eta i}{2})-\tfrac{1+\tau}{2})}{\sigma(i(u_j+\tfrac{\eta i}{2})+\tfrac{1}{2})\sigma(i(u_j+\tfrac{\eta i}{2})+\tfrac{\tau}{2})\sigma(i(u_j+\tfrac{\eta i}{2})-\tfrac{1+\tau}{2})}\\
       &~~~\times \prod_{\gamma=\pm}\prod_{k=1}^{3}\frac{\sigma(i(u_j+\epsilon_k^{\gamma}\alpha_k^{\gamma}i+\tfrac{\eta}{2} i))}{\sigma(i(u_j-\epsilon_k^{\gamma}\alpha_k^{\gamma}i-\tfrac{\eta}{2} i))}(\frac{\sigma(i(u_j-\frac{\eta}{2} i))}{\sigma(i(u_j+\tfrac{\eta}{2} i))})^{2L+1}\\
       &=\prod_{l=1}^{L}\frac{\sigma(i(u_j-u_l-\eta i))\sigma(i(u_j+u_l-\eta i))}{\sigma(i(u_j-u_l+\eta i))\sigma(i(u_j+u_l+\eta i))},\qquad j=1,\cdots L
    \end{aligned}
\end{equation}
where $\sigma(u)$ is the elliptic function defined in Section \ref{s2}. On the gauge side, the exact partition function $\mathcal{I}$ of the 4D supersymmetric gauge theory on $D^2\times T^2$ with Dirichlet and Robin-like boundary conditions was computed via supersymmetry localization \cite{LNP19}. Let $\mathfrak{g}$ be the Lie algebra of a gauge group $G$, and $\mathfrak{h}$ is the Cartan subalgebra of $\mathfrak{g}$. We derive the effective superpotential $W^{4d}_{\text{eff}}(\Phi_{(0)},m)$ through the relation
\begin{equation}\label{2}
    \mathcal{I}\sim \text{exp}\left(\dfrac{1}{\epsilon}W^{4d}_{\text{eff}}(\Phi_{(0)},m)\right)
\end{equation}
where $\epsilon$ is the $\Omega$-background deformation parameter in Nekrasov-Shatashvili (NS) limit, $\Phi_{(0)}\in \mathfrak{h}$ denotes the zero mode of the gauge connection. In our framework, the norm of the root vectors in the Lie algebra $\mathfrak{g}$ (associated with gauge group $G$) is explicitly embedded into the effective superpotential. This leads to the vacuum equation
\begin{equation}\label{120}
\text{exp}\left(\sigma \dfrac{\partial}{\partial (\Phi_{(0)})}W^{4d}_{\text{eff}}(\Phi_{(0)},m)\right)=1
\end{equation}
where $\sigma$ is the disk equivariant parameter. For notational clarity, we henceforth denote $\Phi_i$ by $\{\Phi_{(0)}\}_i$ in subsequent analyses.

We derive four classes of vacuum equations corresponding to Robin-like boundary conditions for classical Lie groups. Specifically, for the $A$-type gauge group, the vacuum equations are 
 \begin{equation}\label{}
\begin{aligned}
-&e^{Ph_{A}}\prod_{j=1}^N\frac{\sigma(\Phi_i-\Phi_j+m_{adj})}{\sigma(\Phi_i-\Phi_j-m_{adj})}=\prod_{a=1}^{N_f}\frac{\sigma (\Phi_i-m_a^{'})}{\sigma (\Phi_i+m_a)}
\end{aligned}
\end{equation}
where
\begin{equation*}
\begin{aligned}
Ph_{A}=&\sum_{j\neq i}^N\frac{-4\pi i(\Phi_i-\Phi_j)m_{adj}}{\tau}+\sum_{a=1}^{N_f}\left[\frac{2\pi i\Phi_i(m_a+m_a^{'})}{\tau}\right.\\
&+ \left.\frac{i\pi(m_a^2-(m_a^{'})^2)-2\pi i(1+\tau)\Phi_i-\pi i(1+\tau)(m_a-m_a^{'})}{\tau}\right]
\end{aligned}
\end{equation*}
For the $B$-type gauge group, the vacuum equations are
\begin{equation}\label{}
\begin{aligned}
    &e^{Ph_{B}}\prod_{i<j}\frac{\sigma(\Phi_{i}\pm \Phi_{j}-m_{adj})}{\sigma(\Phi_{i}\pm \Phi_{j}+m_{adj})} \frac{\sigma(2(\Phi_{i}-m_{adj}))^2}{\sigma(2(\Phi_{i}+m_{adj}))^2}=\prod_{a=1}^{N_f}\frac{\sigma(\Phi_{i}-m_a)}{\sigma(\Phi_{i}+m_a)},\quad i=1,\cdots,N
\end{aligned}
\end{equation}
where
\begin{equation*}
\begin{aligned}
Ph_{B}=&\sum_{i<j}\frac{8i\pi \Phi_{i}m_{adj}-4i\pi(1+\tau) m_{adj}+2\pi i\tau(\Phi_i\pm \Phi_j)}{\tau}\\
&+\sum_{a=1}^{N_f}\dfrac{i\pi(4\Phi_{i}m_a-2\tau\Phi_i)}{\tau}+\frac{16\pi i\Phi_im_{adj}+4\tau \pi i\Phi_i}{\tau}
\end{aligned}
\end{equation*}
For the $C$-type gauge group, the vacuum equations are
\begin{equation}\label{}
\begin{aligned}
    &e^{Ph_{C}} \prod_{i<j}\frac{\sigma(\Phi_{i}\pm \Phi_{j}-m_{adj})}{\sigma(\Phi_{i}\pm \Phi_{j}+m_{adj})} \frac{\sigma(\Phi_{i}-m_{adj})}{\sigma(\Phi_{i}+m_{adj})}=\prod_{a=1}^{N_f}\frac{\sigma(\Phi_{i}-m_a)}{\sigma(\Phi_{i}+m_a)},\quad i=1,\cdots,N
\end{aligned}
\end{equation}
where
\begin{equation*}
\begin{aligned}
Ph_{C}=&\sum_{i<j}\frac{8i\pi \Phi_{i}m_{adj}-4i\pi(1+\tau) m_{adj}+2\pi i\tau(\Phi_i\pm \Phi_j)}{\tau}\\
&+\sum_{a=1}^{N_f}\dfrac{i\pi(4\Phi_{i}m_a-2(1+\tau)\Phi_i+2\Phi_i)}{\tau}+\frac{4\pi i\Phi_im_{adj}+2\tau\pi i\Phi_i}{\tau}
\end{aligned}
\end{equation*}
For the $D$-type gauge theory, the vacuum equations are
\begin{equation}\label{}
\begin{aligned}
    e^{Ph_{D}} \prod_{i<j}\frac{\sigma(\Phi_{i}\pm \Phi_{j}-m_{adj})}{\sigma(\Phi_{i}\pm \Phi_{j}+m_{adj})} =\prod_{a=1}^{N_f}\frac{\sigma(\Phi_{i}-m_a)}{\sigma(\Phi_{i}+m_a)},
    \quad i=1,\cdots,N
\end{aligned}
\end{equation}
where
\begin{equation*}
\begin{aligned}
Ph_{D}=&\sum_{i<j}\frac{8i\pi \Phi_{i}m_{adj}-4i\pi(1+\tau) m_{adj}+2\pi i\tau(\Phi_i\pm \Phi_j)}{\tau}\\
&+\sum_{a=1}^{N_f}\dfrac{i\pi(4\Phi_{i}m_a-2(1+\tau)\Phi_i+2\Phi_i\tau)}{\tau}
\end{aligned}
\end{equation*}

The core findings of this work establish a systematic Bethe/gauge correspondence through the following dualities:

\begin{enumerate}
    \item \textbf{A-type gauge theory}: Exact equivalence between
    \begin{itemize}
        \item $A_{n}$-type vacuum equations and closed-boundary BAEs
        \item Gauge group $SU(n+1)$ theories for (2D/3D/4D) supersymmetric theories, duality with periodic $\text{XXX}$/$\text{XXZ}$/$\text{XYZ}$ spin chains, respectively.
    \end{itemize}
    
    \item \textbf{BCD-type theories}: Universal mapping relating
    \begin{itemize}
        \item $B_n(SO(2n+1))/C_n(Sp(2n))/D_n(SO(2n))$-type vacuum equations to open-boundary BAEs
        \item Orthogonal/Symplectic gauge theories (2D/3D/4D) to $\text{XXX}$/$\text{XXZ}$/$\text{XYZ}$ spin chains with diagonal/general boundary conditions:
        \begin{itemize}[label=\checkmark]
            \item 2D:  $SO(n)/Sp(2n) \leftrightarrow$ open $\text{XXX}$ spin chain
            \item 3D:  $SO(n)/Sp(2n) \leftrightarrow$ open $\text{XXZ}$ spin chain
            \item 4D:  $SO(n)/Sp(2n) \leftrightarrow$ open $\text{XYZ}$ spin chain
        \end{itemize}
    \end{itemize}
\end{enumerate}

Crucially, the gauge group $G$ determines both the vacuum equations parameters and the required boundary conditions matching in (\ref{1}). 

The exact duality correspondence between supersymmetric gauge theories with distinct spacetime dimensions and gauge groups, and quantum spin chains subject to different boundary conditions, remains to be explored in the existing Bethe/gauge correspondence framework. Our methodology systematically establishes these dualities through 2D, 3D, and 4D supersymmetry gauge theories and quantum $\text{XXX}$, $\text{XXZ}$ and $\text{XYZ}$ spin chains with the most general boundary conditions. This framework extends previous results by demonstrating unified correspondence rules across various spacetime dimensions, and with different Lie group types, confirming universal quantum integrability structures underlying supersymmetric gauge dynamics.

The paper is organized as follows. We first introduce the open $\text{XYZ}$ spin chain with general boundary condition and the BAEs in section \ref{s2}. In section \ref{s3}, we give the effective superpotential of different boundary condition. In section \ref{s4}, we calculate the vacuum equations for a given gauge group $G$ and compare these vacuum equations with the BAEs to obtain the Bethe/gauge correspondence. The last section is the conclusions and discussions. As a natural deduction of our approach, we present in Appendix \ref{secA1} the ‌exact duality‌ between supersymmetric gauge theories of $\rm BCD$-type and the $\rm XXZ$ spin chain with generic boundary conditions‌.

\section{The $\text{XYZ}$ spin chain}\label{s2}

\subsection{The Hamiltonian}

In this section, we introduce some results presented in \cite{CYCSW14, XCYW24} about the $\text{XYZ}$ spin chain with periodic boundary condition and open boundary condition. The Hamiltonian of the $\text{XYZ}$ spin chain with periodic boundary condition is
\begin{equation}\label{}
\begin{aligned}
H=&\frac{1}{2}\sum_{j=1}^{L}(J_x\sigma_j^x\sigma^x_{j+1}+J_y\sigma_j^y\sigma^y_{j+1}+J_z\sigma_j^z\sigma^z_{j+1})
\end{aligned}
\end{equation}
where $\sigma_j^{\alpha}(\alpha=x, y, z)$ is the Pauli matrix on the j-$\text{th}$ site along the $\alpha$ direction and {red}$L$ is the number of sites. The anisotropic couplings in the bulk are
\begin{equation}\label{6}
J_x=\frac{e^{i\pi \eta}\sigma(\eta+\tfrac{\tau}{2})}{\sigma(\tfrac{\tau}{2})},\qquad J_y=\frac{e^{i\pi \eta}\sigma(\eta+\tfrac{1+\tau}{2})}{\sigma(\tfrac{1+\tau}{2})},\qquad J_z=\frac{\sigma(\eta+\tfrac{1}{2})}{\sigma(\tfrac{1}{2})}
\end{equation}
where $\eta$ is the crossing parameter and $\tau$ is the modulus parameter. The periodic boundary condition means
$$\sigma_{L+1}^x=\sigma_1^x,\quad \sigma_{L+1}^y=\sigma_1^y, \quad \sigma_{L+1}^z=\sigma_1^z$$

The authors studied the thermodynamic limit of the anisotropic $\text{XYZ}$ spin chain with non-diagonal integrable open boundary conditions in \cite{XCYW24}. The Hamiltonian of the $\text{XYZ}$ spin chain with generic boundary condition \cite{WYC15} reads
\begin{equation}\label{5}
\begin{aligned}
H=&\sum_{j=1}^{L-1}(J_x\sigma_j^x\sigma^x_{j+1}+J_y\sigma_j^y\sigma^y_{j+1}+J_z\sigma_j^z\sigma^z_{j+1})+h_x^{-}\sigma_1^x+h_y^{-}\sigma_1^y+h_z^{-}\sigma_1^z\\
&+h_x^{+}\sigma_N^x+h_y^{+}\sigma_N^y+h_z^{+}\sigma_N^z
\end{aligned}
\end{equation}
The boundary magnetic fields are
\begin{equation}
\begin{aligned}
&h_x^{\mp}=\pm e^{-i\pi(\sum_{l=1}^3\alpha_l^{\mp}-\tfrac{\tau}{2})}\frac{\sigma(\eta)}{\sigma(\tfrac{\tau}{2})}\prod_{l=1}^3\frac{\sigma(\alpha_l^{\mp}-\tfrac{\tau}{2})}{\sigma(\alpha_l^{\mp})}\\
&h_y^{\mp}=\pm e^{-i\pi(\sum_{l=1}^3\alpha_l^{\mp}-\tfrac{1+\tau}{2})}\frac{\sigma(\eta)}{\sigma(\tfrac{1+\tau}{2})}\prod_{l=1}^3\frac{\sigma(\alpha_l^{\mp}-\tfrac{1+\tau}{2})}{\sigma(\alpha_l^{\mp})}\\
&h_z^{\mp}=\pm \frac{\sigma(\eta)}{\sigma(\tfrac{1}{2})}\prod_{l=1}^3\frac{\sigma(\alpha_l^{\mp}-\tfrac{1}{2})}{\sigma(\alpha_l^{\mp})}\\
\end{aligned}
\end{equation}
where $\sigma(u)$ is defined by elliptic function (which we will introduce in the following) and $\{\alpha_l^{\mp}|l=1,2,3\}$ are free boundary parameters which specify the strengths of boundary magnetic fields.

The elliptic functions are defined as 
\begin{equation*}
\begin{aligned}
    &\theta_{a,b}(u,\tau)=\sum_m e^{i\pi (m+a)^2 \tau+2i\pi (m+a)(u+b)}\\
    &\sigma(u)=\theta_{1/2,1/2}(u,\tau)=\sum_m e^{i\pi (m+\tfrac{1}{2})^2 \tau+2i\pi (m+\tfrac{1}{2})(u+\tfrac{1}{2})}
\end{aligned}
\end{equation*}
where $u$ is the spectral parameter, $a, b$ are the rational numbers and $\tau$ is the modulus parameters with $Im(\tau)>0$. For our analysis, we employ the Jacobi triple product identity in the following form \cite{FK01}. Let $z \in \mathbb{C}\setminus\{0\}$ and $x \in \mathbb{C}$ with $|x| < 1$, then the identity reads:
   \begin{equation}
        \prod_{n=1}^{\infty}(1-x^{2n})(1+x^{2n-1}z)(1+\frac{x^{2n-1}}{z})=\sum_{n=-\infty}^{\infty}x^{n^2}z^n
    \end{equation}
Then one has
\begin{equation}
\begin{aligned}
    \sigma(u)&=\sum_m e^{i\pi (m+\tfrac{1}{2})^2 \tau+2i\pi (m+\tfrac{1}{2})(u+\tfrac{1}{2})}\\
    &=iq^{\tfrac{1}{8}}e^{\pi iu}\prod_{m=1}^{\infty}(1-q^m)(1-q^me^{2\pi iu})(1-q^{m-1}e^{-2\pi iu})
\end{aligned}
\end{equation}
where $q=e^{2i\pi \tau}$.

The integrability of the model (\ref{5}) is associated with the eight-vertex $R$-matrix $R(u)\in End(\mathbb{C}^2\times \mathbb{C}^2)$
\begin{equation}
R(u)=\begin{pmatrix}
a(u) &  &  & d(u)\\
 & b(u) & c(u) & \\
 & c(u) & b(u) & \\
d(u) & & & a(u)\\
\end{pmatrix}
\end{equation}
The non-vanishing matrix entries are
\begin{equation*}
\begin{aligned}
&a(u)=\frac{\theta_{0,\tfrac{1}{2}}(u,2\tau)\theta_{\tfrac{1}{2},\tfrac{1}{2}}(u+\eta,2\tau)}{\theta_{0,\tfrac{1}{2}}(0,2\tau)\theta_{\tfrac{1}{2},\tfrac{1}{2}}(\eta,2\tau)},\quad b(u)=\frac{\theta_{\tfrac{1}{2},\tfrac{1}{2}}(u,2\tau)\theta_{0,\tfrac{1}{2}}(u+\eta,2\tau)}{\theta_{0,\tfrac{1}{2}}(0,2\tau)\theta_{\tfrac{1}{2},\tfrac{1}{2}}(\eta,2\tau)}\\
&c(u)=\frac{\theta_{0,\tfrac{1}{2}}(u,2\tau)\theta_{0,\tfrac{1}{2}}(u+\eta,2\tau)}{\theta_{0,\tfrac{1}{2}}(0,2\tau)\theta_{0,\tfrac{1}{2}}(\eta,2\tau)},\quad d(u)=\frac{\theta_{\tfrac{1}{2},\tfrac{1}{2}}(u,2\tau)\theta_{\tfrac{1}{2},\tfrac{1}{2}}(u+\eta,2\tau)}{\theta_{0,\tfrac{1}{2}}(0,2\tau)\theta_{0,\tfrac{1}{2}}(\eta,2\tau)}\\
\end{aligned}
\end{equation*}
The $R$-matrix satisfies the quantum Yang-Baxter equation
\begin{equation}
R_{12}(u_1-u_2)R_{13}(u_1-u_3)R_{23}(u_2-u_3)=R_{23}(u_2-u_3)R_{13}(u_1-u_3)R_{12}(u_1-u_2)
\end{equation}
and other properties, see \cite{XCYW24} for details. The open boundary condition is characterized by a pair of reflection matrices $K^-(u)$ and $K^+(u)$. The former satisfies the reflection equation (RE)
\begin{equation*}
\begin{aligned}
R_{12}(u_1-u_2)&K_1^-(u_1)R_{21}(u_2+u_1)K_2^-(u_2)\\
=&K_2^-(u_2)R_{12}(u_2+u_1)K_1^-(u_1)R_{21}(u_1-u_2)
\end{aligned}
\end{equation*}
and the latter satisfies the dual RE
\begin{equation*}
\begin{aligned}
R_{12}(u_1-u_2)&K_1^+(u_1)R_{21}(-u_1-u_2-2)K_2^+(u_2)\\
=&K_2^+(u_2)R_{12}(-u_1-u_2-2)K_1^-(u_1)R_{21}(u_1-u_2)
\end{aligned}
\end{equation*}
where $K_{-}^{1}(u):=K_{-}(u)\otimes id_{V_{2}}$ and $K_{-}^{2}(u):=id_{V_{1}} \otimes K_{-}(u)$. Hence the transfer matrix $t(u)$ of the close and open $\text{XYZ}$ spin chain are given by
\begin{equation}
\begin{aligned}
t(u)&=tr_0T_0(u)\\
t(u)&=tr_0{K_0^+(u)T_0(u)K_0^-(u)\hat{T}_0(u)}
\end{aligned}
\end{equation}
where $T_0(u)$ and $\hat{T}_0(u)$ are the monodromy matrices
\begin{equation*}
\begin{aligned}
&T_0(u)=R_{0L}(u-\theta_N)\cdots R_{01}(u-\theta_1)\\
&\hat{T}_0(u)=R_{10}(u+\theta_1)\cdots R_{N0}(u+\theta_N)
\end{aligned}
\end{equation*}
Here $\theta_j, j=1,\cdots, N$ are the inhomogeneity parameters.

The most general solutions of the RE and dual RE are
\begin{equation}
K^-(u)=\frac{\sigma(2u)}{2\sigma(u)}\Big\{id+\frac{c_x^-\sigma(u)e^{-i\pi u}}{\sigma(u+\tfrac{\tau}{2})}\sigma^x+\frac{c_y^-\sigma(u)e^{-i\pi u}}{\sigma(u+\tfrac{1+\tau}{2})}\sigma^y+\frac{c_z^-\sigma(u)}{\sigma(u+\tfrac{1}{2})}\sigma^z\Big\}
\end{equation}
\begin{equation}
K^+(u)=K^-(-u-\eta)\Big|_{c_l^-\rightarrow c_l^+}
\end{equation}
where the constants $\{c_{\alpha}^\mp|\alpha=x,y,z\}$ are expressed in terms of boundary parameters $\{\alpha_l^\mp|l=1,2,3\}$ as following
\begin{equation}
\begin{aligned}
&c_x^\mp=e^{-i\pi(\sum_{l=1}\alpha_l^{\mp}-\frac{\tau}{2})}\prod_{l=1}^3\frac{\sigma(\alpha_l^\mp-\frac{\tau}{2})}{\sigma(\alpha_l^\mp)}\\
&c_y^\mp=e^{-i\pi(\sum_{l=1}\alpha_l^{\mp}-\frac{1+\tau}{2})}\prod_{l=1}^3\frac{\sigma(\alpha_l^\mp-\frac{1+\tau}{2})}{\sigma(\alpha_l^\mp)}\\
&c_z^\mp=\prod_{l=1}^3\frac{\sigma(\alpha_l^\mp-\frac{1}{2})}{\sigma(\alpha_l^\mp)}\\
\end{aligned}
\end{equation}

It is easy to see that the setting $c_x^{-}=c_y^{-}=0$ derives $K^-(u)$ is diagonal, i.e, 
$$e^{-i\pi(\sum_{l=1}\alpha_l^{\mp}-\frac{\tau}{2})}\prod_{l=1}^3\frac{\sigma(\alpha_l^\mp-\frac{\tau}{2})}{\sigma(\alpha_l^\mp)}=0$$
$$e^{-i\pi(\sum_{l=1}\alpha_l^{\mp}-\frac{1+\tau}{2})}\prod_{l=1}^3\frac{\sigma(\alpha_l^\mp-\frac{1+\tau}{2})}{\sigma(\alpha_l^\mp)}=0$$
So one obtains $\{\alpha_l^\mp|=1,2,3\}\rightarrow -i\infty$.

\subsection{The Bethe Ansatz equation}
The the BAEs of closed $\text{XYZ}$ spin chain are obtained by $T-Q$ relation
\begin{equation}\label{71}
e^{4\pi il_1u_j+2i\phi}\frac{\sigma^L(u_j+\eta)}{\sigma^L(u_j)}=-\prod_{k=1}^{\bar{M}}\frac{\sigma(u_j-u_k+\eta)}{\sigma(u_j-u_k-\eta)},\qquad j=1,\cdots, \bar{M}
\end{equation}
where $M_1$ is the number of Bethe roots, $\bar{M}=M_1+m$ is a non-negative integer and $l_1$, $m_1$ are a integral. The relation is $(\frac{L}{2}-m-M_1)\eta=l_1\tau+m_1$. See \cite{CYCSW14} for details.

For real crossing parameter $\eta$, the BAEs of open $\text{XYZ}$ spin chain are obtained by 
\begin{equation}\label{7}
    \begin{aligned}
       &-\frac{\sigma(i(2u_j-\eta i))}{\sigma(i(2u_j+\eta i))}\prod_{\gamma=\pm}\prod_{k=1}^{3}\frac{\sigma(i(u_j+\epsilon_k^{\gamma}\alpha_k^{\gamma}i+\tfrac{\eta}{2} i))}{\sigma(i(u_j-\epsilon_k^{\gamma}\alpha_k^{\gamma}i-\tfrac{\eta}{2} i))}\Big(\frac{\sigma(i(u_j-\frac{\eta}{2} i))}{\sigma(i(u_j+\tfrac{\eta}{2} i))}\Big)^{2L}\\
       &~~=\prod_{l=1}^{L}\frac{\sigma(i(u_j-u_l-\eta i))\sigma(i(u_j+u_l-\eta i))}{\sigma(i(u_j-u_l+\eta i))\sigma(i(u_j+u_l+\eta i))},\qquad j=1,\cdots L
    \end{aligned}
\end{equation}
where $\alpha_k^{\pm}$ are boundary parameters and
\begin{equation}
    \prod_{\gamma=\pm}\prod_{j=1}^{3}\epsilon_j^{\gamma}=-1,\qquad \epsilon_j^{\gamma}=\pm1
\end{equation}
Furthermore, we utilize the following identities in elliptic functions throughout the derivation process
\begin{equation}
    \sigma(2u)=\frac{2\sigma(u)\sigma(u+\tfrac{1}{2})\sigma(u+\tfrac{\tau}{2})\sigma(u-\tfrac{1+\tau}{2})}{\sigma(\tfrac{1}{2})\sigma(\tfrac{\tau}{2})\sigma(-\tfrac{1+\tau}{2})}
\end{equation}

Hence we get the BAEs equivalently 
\begin{equation}\label{8}
    \begin{aligned}
       &-\frac{\sigma(i(u_j-\tfrac{\eta i}{2})+\tfrac{1}{2})\sigma(i(u_j-\tfrac{\eta i}{2})+\tfrac{\tau}{2})\sigma(i(u_j-\tfrac{\eta i}{2})-\tfrac{1+\tau}{2})}{\sigma(i(u_j+\tfrac{\eta i}{2})+\tfrac{1}{2})\sigma(i(u_j+\tfrac{\eta i}{2})+\tfrac{\tau}{2})\sigma(i(u_j+\tfrac{\eta i}{2})-\tfrac{1+\tau}{2})}\\
       &\times \prod_{\gamma=\pm}\prod_{k=1}^{3}\frac{\sigma(i(u_j+\epsilon_k^{\gamma}\alpha_k^{\gamma}i+\tfrac{\eta}{2} i))}{\sigma(i(u_j-\epsilon_k^{\gamma}\alpha_k^{\gamma}i-\tfrac{\eta}{2} i))}(\frac{\sigma(i(u_j-\frac{\eta}{2} i))}{\sigma(i(u_j+\tfrac{\eta}{2} i))})^{2L+1}\\
       &=\prod_{l=1}^{L}\frac{\sigma(i(u_j-u_l-\eta i))\sigma(i(u_j+u_l-\eta i))}{\sigma(i(u_j-u_l+\eta i))\sigma(i(u_j+u_l+\eta i))},\qquad j=1,\cdots L
    \end{aligned}
\end{equation}

In the $\rm XXZ$ limit by setting $J_x=J_y$, i.e., 
$$\frac{e^{i\pi \eta}\sigma(\eta+\tfrac{\tau}{2})}{\sigma(\tfrac{\tau}{2})}=\frac{e^{i\pi \eta}\sigma(\eta+\tfrac{1+\tau}{2})}{\sigma(\tfrac{1+\tau}{2})}$$
we get $\tau\rightarrow i\infty$ $(q\rightarrow 0)$. Then we have
$$J_x\rightarrow 1, \qquad J_y\rightarrow 1, \qquad J_z\rightarrow {\rm cos}\pi \eta$$
$$d(u)\rightarrow 0$$
$$\sigma(u) \sim -2q^{\tfrac{1}{8}}{\rm sin}(\pi u),\qquad \sigma(u\pm \tfrac{1}{2})\sim \pm2q^{\tfrac{1}{8}}{\rm cos}(\pi u)$$
$$\sigma(u\pm \tfrac{\tau}{2}) \sim \mp i\infty,\qquad \sigma(u\pm \tfrac{1+\tau}{2})\sim \mp\infty$$
By the way
$$\frac{\sigma(u)e^{-\pi iu}}{\sigma(u+\tfrac{\tau}{2})}\sim \frac{1-e^{-2\pi iu}}{q^{1/4}e^{\pi iu}-q^{-1/4}e^{-\pi iu}}\rightarrow 0$$
$$\frac{\sigma(u)e^{-\pi iu}}{\sigma(u+\tfrac{1+\tau}{2})}\sim -i\frac{1-e^{-2\pi iu}}{q^{1/4}e^{\pi iu}-q^{-1/4}e^{-\pi iu}}\rightarrow 0$$
we have a off-diagonal matrix
\begin{equation}
\begin{aligned}
K^-(u)&=\begin{pmatrix}
&{\rm cos}(\pi u)+{\rm sin}(\pi u)c_x^- & \chi(u) (c_x^--c_y^-)\\
&\chi (u) (c_x^-+c_y^-) & {\rm cos}(\pi u)-{\rm sin}(\pi u)c_x^-
\end{pmatrix}\\
\end{aligned}
\end{equation}
in which $ \chi (u)= \frac{{\rm cos}(\pi u)(1-e^{-2\pi iu})}{q^{1/4}e^{\pi iu}(1-e^{-2\pi i(u+\tau/2)})}$. So we get the $\rm XXZ$ spin chain with the most general off-diagonal boundary condition. The BAEs are degenerated to
\begin{equation}\label{100}
    \begin{aligned}
       &-\frac{{\rm sin}(\pi(2iu_j+\eta))}{{\rm sin}(\pi(2iu_j-\eta))}\prod_{\gamma=\pm}\prod_{k=1}^{3}\frac{{\rm sin}(\pi(iu_j-\epsilon_k^{\gamma}\alpha_k^{\gamma}-\tfrac{\eta}{2}))}{{\rm sin}(\pi(iu_j+\epsilon_k^{\gamma}\alpha_k^{\gamma}+\tfrac{\eta}{2}))}\Big(\frac{{\rm sin}(\pi(iu_j+\frac{\eta}{2}))}{{\rm sin}(\pi(iu_j-\tfrac{\eta}{2}))}\Big)^{2L}\\
       &~~=\prod_{l=1}^{L}\frac{{\rm sin}(\pi(iu_j-iu_l+\eta)){\rm sin}(\pi(iu_j+iu_l+\eta))}{{\rm sin}(\pi(iu_j-iu_l-\eta)){\rm sin}(\pi(iu_j+iu_l-\eta ))},\qquad j=1,\cdots L
    \end{aligned}
\end{equation}
One can further take the limit $\eta\rightarrow 0$ with the spectral parameter rescaled by $u\rightarrow u\eta$ to go to the $\rm XXX$ limit. From the previously used setup, one can only derive the explicit supersymmetric gauge theories duality with the diagonal boundary conditions in the $\rm XXZ$ and $\rm XXX$ cases. See the Appendix \ref{secA1}.

\section{The effective superpotential}\label{s3}

In \cite{LNP19}, the authors studied 4D $\mathcal{N}=1$ gauge theories with $R$-symmetry on a hemisphere times torus geometry. They applied localization techniques to compute the exact partition function via a cohomological reformulation of supersymmetry transformations. In the remainder of this subsection, we introduce the effective superpotential derived from the partition function.  

\subsection{The 1-loop determinants}

We start from useful definitions and properties of some special functions used in the gauge theory. 

The (infinite) $q$-factorial is defined by
\begin{equation}
    (x;q)_\infty=\prod_{k>0}(1-q^kx), \quad |q|<1
\end{equation}
Using the representation
$$(x;q)_\infty=e^{-\text{Li}_2(x;q)},\qquad \text{Li}_2(x;q)=\sum_{k\ge 1}\frac{x^k}{k(1-q^k)}$$
it  can be extended to the domain $|q|>1$ by means of 
$$(qx;q)_\infty\rightarrow \frac{1}{(x;q^{-1})}_\infty$$
The double (infinite) $q$-factorial is defined by
\begin{equation*}
\begin{aligned}
    &(x;p,q)_{\infty}=\prod_{j,k\ge 0}(1-p^jq^kx)=e^{-\text{Li}_3(x;p,q)},\qquad |p|,|q|>1\\
    &\text{Li}_3(x;p.q)=\sum_{j,k\ge 1}\frac{x^k}{k(1-p^j)(1-q^k)}
\end{aligned}
\end{equation*}
The elliptic Gamma function is defined by 
\begin{equation*}
    \Gamma(x;p,q)=\dfrac{(pqx^{-1};p,q)_{\infty}}{(x;p,q)_{\infty}}=\prod_{m\ge 0}\prod_{n\ge 0}\dfrac{1-q^{m+1}q^{n+1}x^{-1}}{1-p^{m}q^{n}x}
\end{equation*}

In the main text we will often using the alternative notation
\begin{equation*}
\Gamma(x; q, p)=\Gamma(u; \tau, \sigma) ,\qquad x=e^{2\pi iu},\quad q=e^{2\pi i\tau},
\quad p=e^{2\pi i \sigma},
\end{equation*}
The Jacobi Theta function is defined by
\begin{equation*}
\Theta(x;q)=\Gamma(x;q)^{-1}\Gamma(qx^{-1};q)^{-1}=\prod_{k\ge 0}(1-q^{k}x)(1-q^{k+1}x^{-1})
\end{equation*}
$P_{3}$ is a cubic Bernoulli polynomial $B_{33}$ up to a constant
\begin{equation}
\begin{aligned}
    P_{3}(X)=&\dfrac{X^{3}}{\tau \sigma}-\dfrac{3(1+\tau+\sigma)X^{2}}{2\tau\sigma}
    +\dfrac{1+\tau^{2}\sigma^{2}+3(\tau+\sigma+\tau\sigma)}{2\tau\sigma}X \\
    &-\dfrac{(1+\tau+\sigma)(\tau+\sigma+\tau \sigma)}{4\tau\sigma}
    -\dfrac{1-\tau^{2}+\tau^{4}}{24\sigma(\tau+\tau^{2})}
\end{aligned}    
\end{equation}

The 1-loop determinants of 4D $\mathcal{N} = 1$ vector and chiral multiplets on $D^2\times T^2$, which constitute the building blocks of integral expressions of gauge theory partition functions. For a vector multiplet of a gauge group $G$ with its Cartan subalgebra $\mathfrak{h}$, the 1-loop determinants can be written as \cite{LNP19}
\begin{equation}\label{50}
    \mathcal{Z}_{1-loop}^{\text{vec}}(\Phi_{(0)})=\left[\dfrac{e^{-\frac{i\pi}{3}P_{3}(0)}}{\text{Res}_{u=0}\Gamma(u;\tau,\sigma)}\right]^{rk(G)}\text{det}_{ad}\left[\dfrac{e^{-\frac{i\pi}{3}P_{3}(\Phi_{(0)})}}{\Gamma(\Phi_{(0)};\tau,\sigma)}\right]
\end{equation}
where $\Phi_{(0)}\in \mathfrak{h}$ is the zero mode of the gauge connection, $P_{3}(u)$ is a cubic polynomial arising from regularization, $\Gamma(u;\tau,\sigma)$ is the elliptic Gamma function with parameters $\tau,\sigma$ associated with the torus modulus and disk equivariant parameters respectively. Similarly,  for a chiral multilplet of $R$-charge $r$ in a representation $\mathcal{R}$ of the gauge group $G$, the one-loop determinants are
\begin{equation}\label{51}
    \mathcal{Z}_{1-loop}^{\text{chi(D)}}(\Phi_{(0)})=\text{det}_{\mathcal{R}}\left[\dfrac{e^{-\frac{i\pi}{3}P_{3}((1-\frac{r}{2})\sigma-\Phi_{(0)})}}{\Gamma((1-\frac{r}{2})\sigma-\Phi_{(0)};\tau,\sigma)}\right]
\end{equation}
for Dirichlet boundary conditions, and 
\begin{equation}\label{52}
    \mathcal{Z}_{1-loop}^{\text{chi(R)}}(\Phi_{(0)})=\text{det}_{\mathcal{R}}\left[e^{\frac{i\pi}{3}P_{3}(\frac{r}{2}\sigma+\Phi_{(0)}}\Gamma(\frac{r}{2}\sigma+\Phi_{(0)};\tau,\sigma)\right]
\end{equation}
for Robin-like boundary. The ratio of these above results is a Jocabi Theta function (up to the exponential of a quadratic polynomial)
\begin{equation}
    \dfrac{\mathcal{Z}_{1-loop}^{\text{chi(D)}}(\Phi_{(0)})}{\mathcal{Z}_{1-loop}^\text{chi(R)}\Phi_{(0)}}=\text{det}_{\mathcal{R}}\left[e^{-i\pi P_{2}(\frac{r}{2}\sigma+\Phi_{(0)})}\Theta(\frac{r}{2}\sigma+\phi_{(0)};\sigma)\right]
\end{equation}

\subsection{The effective superpotential} 

In 3D, the effective twisted superpotential $W^{3d}_{\text{eff}}(\sigma,m)$ is obtained by the partition function $\mathcal{I}$ 
\begin{equation}\label{9}
    \mathcal{I}^{3d}\sim \text{exp}\left(\dfrac{1}{\epsilon}W^{3d}_{\text{eff}}(\sigma,m)\right)
\end{equation}
where $\epsilon$ is the $\Omega$-deformation parameter in NS limit \cite{DZ23a, KZ21}. Similarly, in 4D, the relation of the effective superpotential $W^{4d}_{\text{eff}}(\Phi_{(0)},m)$ and the partition function $\mathcal{I}^{4d}$ is given by (\ref{2}). Actually, we need to calculate one loop determinants $\mathcal{Z}$ in the partition function to get effective twisted superpotential. 

We apply to the method to 4D parallelly. In this article, we consider the representation of gauge group $G$
\begin{equation}\label{10}
    \mathcal{R}=V\otimes V^{*}\oplus V\otimes F
\end{equation}
which corresponds to the theory with the $H^{\text{max}}=U(L)$ global symmetry group. Here $V=\mathbf{C}^{N}$ is the $N$-dimensional fundamental representation of $G$, $\mathcal{F}\approx \mathbf{C}^{N_f}$ is the $L$-dimensional fundamental representations of the $U(L)$. With the identity
\begin{equation*}
    \text{log}(\text{det}Z)=\text{tr}(\text{log}Z)
\end{equation*}
we get the effective superpotential of Robin boundary condition
\begin{equation}\label{11}
    \begin{aligned}
     W^{4d,\text{R}}_{\text{eff}}(\Phi_{(0)})=&-\sum_{\alpha}\dfrac{i\pi(\tfrac{2}{\alpha^2}\alpha \cdot \Phi_{(0)})^{3}}{3\tau \sigma}+\sum_{\alpha}\dfrac{i\pi(1+\tau)(\tfrac{2}{\alpha^2}\alpha \cdot\Phi_{(0)})^{2}}{2\tau\sigma}\\
     &-\sum_{\alpha}\dfrac{i\pi (1+3\tau)(\tfrac{2}{\alpha^2}\alpha \cdot \Phi_{(0)})}{6\tau \sigma}\\
     &+\sum_{\alpha}\left(\text{Li}_3(e^{2\pi i(\tfrac{2}{\alpha^2}\alpha \cdot \Phi_{(0)})};p.q)-\text{Li}_3(pqe^{-2\pi i(\tfrac{2}{\alpha^2}\alpha \cdot \Phi_{(0)})};p,q)\right)\\
     &+\sum_{\omega}\sum_{f}\dfrac{i\pi (\omega \cdot \Phi_{(0)}+m_f)^{3}}{3\tau \sigma}-\sum_{\omega}\sum_f\dfrac{i\pi (1+\tau)(\omega \cdot \Phi_{(0)}+m_f)^{2}}{2\tau \sigma}\\
     &+\sum_{\omega}\sum_f\dfrac{i\pi(1+3\tau)(\omega \cdot \Phi_{(0)}+m_f)}{6 \tau \sigma}\\
     &+\sum_{\omega}\sum_{f}\left(\text{Li}_3(e^{2\pi i(\omega \cdot \Phi_{(0)}+m_a^f)};p.q)\right.\\
     &~~~~~~~~~~~~~~-\left.\text{Li}_3(pqe^{-2\pi i(\omega \cdot \Phi_{(0)}+m_a^f)};p;q)\right)\\
\end{aligned}
\end{equation}
Equivalently, one can express $W^{4d,\text{R}}_{\text{eff}}(\Phi_{(0)})$ in terms of $Li_2$ functions
\begin{equation}
\begin{aligned}
     W^{4d,\text{R}}_{\text{eff}}(\Phi_{(0)})=&-\sum_{\alpha}\dfrac{i\pi(\tfrac{2}{\alpha^2}\alpha \cdot \Phi_{(0)})^{3}}{3\tau \sigma}+\sum_{\alpha}\dfrac{i\pi(1+\tau)(\tfrac{2}{\alpha^2}\alpha \cdot\Phi_{(0)})^{2}}{2\tau\sigma}\\
     &-\sum_{\alpha}\dfrac{i\pi (1+3\tau)(\tfrac{2}{\alpha^2}\alpha \cdot \Phi_{(0)})}{6\tau \sigma}\\
     &+\dfrac{1}{2\pi i\sigma}\sum_{\alpha}\sum_{n= 1}^{\infty}\left[\text{Li}_{2}(p^{n}e^{2\pi i 
     (\tfrac{2}{\alpha^2}\alpha \cdot \Phi_{(0)})})-\text{Li}_{2}(p^{n-1}e^{-2\pi i(\tfrac{2}{\alpha^2}\alpha \cdot \Phi_{(0)})})\right]\\    
     &+\sum_{\omega}\sum_{f}\dfrac{i\pi (\omega \cdot \Phi_{(0)}+m_f)^{3}}{3\tau \sigma}-\sum_{\omega}\sum_f\dfrac{i\pi (1+\tau)(\omega \cdot \Phi_{(0)}+m_f)^{2}}{2\tau \sigma}\\
     &+\sum_{\omega}\sum_f\dfrac{i\pi(1+3\tau)(\omega \cdot \Phi_{(0)}+m_f)}{6 \tau \sigma}\\
     &+\frac{1}{2\pi i\sigma}\sum_{\omega}\sum_f\sum_{n=1}^{\infty}\left[\text{Li}_2(p^ne^{-2\pi i(\omega \cdot \Phi_{(0)}+m_f)}) \right. \\ 
     &~~~~~~~~~~~~~~~~~~~~~~~~~~\left.{-\text{Li}_2(p^{n-1}e^{2\pi i(\omega \cdot \Phi_{(0)}+m_f)}}\right] \\
\end{aligned}
\end{equation}
where $\alpha$ is the root of the Lie algebra. 

We investigate 4D $\mathcal{N}=1$ supersymmetric gauge theories with generic matter content. The matter sector comprises chiral multiplets, while the gauge field are encoded in vector multiplets. The coefficient $\frac{2}{\alpha^2}$ appearing in the vector multiplet sector is associated with the fundamental representation of the gauge group $G$. 
Here, we establish naturally realizes the Bethe/gauge correspondence with the coefficient $\frac{2}{\alpha^2}$. The vacuum equation 
\begin{equation}\label{12}
\exp\left(\sigma \dfrac{\delta}{\delta \Phi}W^{\rm 4d}_{\rm eff}(\Phi,m)\right) = 1
\end{equation}
is derived from the 3D $\mathcal{N}=2$ gauge theory vacuum configuration on $D^2 \times S^1$. 

Please note that, for the 3D $\mathcal{N}=2$ counterparts analysed in \cite{DZ23a}, the modified coefficient is $\frac{4}{\alpha^2}$ employed in computing the effective superpotential. Of course, in $4D$ case, we can use that setup to get the gauge/Bethe duality. The advantage for convenience of $\frac{2}{\alpha^2}$ over the 3D parameter $\frac{4}{\alpha^2}$ manifests in the elimination of square-root operations during duality transformations, as demonstrated by the streamlined derivation of effective superpotential $W$ and vacuum equation (\ref{12}).

\subsection{Lower dimensional limits}

The dimensional degeneration of 4D gauge theories from $D^2 \times T^2$ to $D^2 \times S^1$ and $D^2$ constitutes a natural geometric reduction framework. As demonstrated in ‌\cite{LNP19}, the 1-loop determinants for dimensionally reduced theories on $D^2 \times S^1$ ‌\cite{YS20} and $D^2$ ‌\cite{FKNO15,HR13} can be systematically derived from the full 4D results through appropriate limit procedures. These simply amount to discarding either one or two towers of KK modes. The determinants in (\ref{51}) and (\ref{52}) reduce to (up to exponentials of quadratic polynomials)
\begin{equation}\label{53}
    \mathcal{Z}_{1-loop}^{\text{vec}}(\Phi_{(0)})\rightarrow \left[\dfrac{1}{\text{Res}_{u=0}\Gamma(u;\sigma)}\right]^{rk(G)}\text{det}_{ad}\left[\frac{1}{\Gamma(\Phi_{(0)};\sigma)}\right]
\end{equation}
\begin{equation}\label{54}
    \mathcal{Z}_{1-loop}^{\text{chi(R)}}(\Phi_{(0)})\rightarrow \text{det}_{\mathcal{R}}\left[\Gamma(\frac{r}{2}\sigma+\Phi_{(0)};\sigma)\right]
\end{equation}
and
\begin{equation}\label{55}
    \mathcal{Z}_{1-loop}^{\text{chi(D)}}(\Phi_{(0)})\rightarrow \text{det}_{\mathcal{R}}\left[\dfrac{1}{\Gamma((1-\frac{r}{2})\sigma-\Phi_{(0)};\sigma)}\right]
\end{equation}
where $\Gamma(u;\sigma)=\prod_{n\ge 0}\frac{1}{1-e^{2\pi i(u+n\sigma)}}$ is the inverse of the q-factorial and hence proportional to the $q$-Gamma function. These results coincide with those in \cite{YS20} for chiral multiplets on $D^2 \times S^1$ with Dirichlet or Neumann boundary condition, respectively.

 The one-loop determinant of the 3D $\mathcal{N} = 2$ vector multiplet:
\begin{equation}\label{56}
\prod_{\alpha\in \Delta}e^{\frac{1}{8\beta_{2}}(\beta l \alpha\cdot \rho)^{2}}(e^{i{\beta l}\alpha\cdot \rho};q^{2})_{\infty},\qquad q=e^{-\beta_{2}}
\end{equation}
where the set of the roots of $\text{Lie}(G)$ is denoted by $\Delta$, the symbol $(a;q)_\infty=\prod_{n=0}^{\infty}(1-aq^n)$. The parameter $\beta_{1}$ is the fugacity of the rotation along $S^{1}$, $\beta_{2}$ is the $U(1)_{\mathcal{R}}$ charge fugacity, $\beta l=(\beta_{1}+\beta_{2})l$ is the circumference of $S^{1}$.  The one-loop determinant of the chiral multiplet with Neumann boundary condition is
\begin{equation}\label{57}
    \mathcal{Z}_{\text{chi}}^{\text{Neu}}=\prod_{w\in \mathcal{R}}e^{\mathcal{E}(i{\beta l}w\cdot \rho+r\beta_{2}+im)}(e^{-i{\beta l}w\cdot \rho-im}q^{r};q^{2})_{\infty}^{-1}
\end{equation}
where the weight's set of the corresponding representation is denoted by $\mathcal{R}$, the $R$-charge of the scalar in the chiral multiplet in $r$, and
\begin{equation*}
    \mathcal{E}(x)=\dfrac{1}{8\beta_{2}}x^{2}-\dfrac{1}{4}x+\dfrac{\beta_{2}}{12}
\end{equation*}
The one-loop contribution of chiral multiplet with Dirichlet boundary condition is 
\begin{equation}\label{58}
\begin{aligned}
\mathcal{Z}_{\text{chi}}^{\text{Dir}}=\prod_{w\in \mathcal{R}}e^{\mathcal{E}(-i{\beta l}w\cdot \rho+(2-r)\beta_{2}-im)}(e^{i{\beta l}w\cdot \rho+im}q^{2-r};q^{2})_{\infty}
\end{aligned}
\end{equation}
In the previous article \cite{DZ23a,DZ23c}, we used the similar method to get the Bethe/gauge correspondence for the open $\text{XXZ}$ spin chain with diagonal boundary condition and the 3D $D^2\times S^1$ gauge theory given with the formula (\ref{56}) and (\ref{57}).

By comparing the BAEs (\ref{100}) and the vacuum equation in \cite{DZ23a} and \cite{DZ23c}, one can establish the duality between the open $\rm XXZ$ spin chain with the most general boundary condition and the 3D $D^2\times S^1$ gauge theory.

\section{Bethe/Gauge correspondence}\label{s4}

In this section, we establish the explicit correspondence between the vacuum equations of A/BCD-type gauge groups and the BAEs (\ref{71})/(\ref{8}) defined in Section \ref{s1}. The 3D case was previously demonstrated in \cite{DZ23a,DZ23c}, with the current 4d results constituting a natural extension of our 3D methodology.

\subsection{$A$-type gauge theory}

Let Euclidean space $\mathbf{E}$ to be the $(N-1)$-dimensional subspace of $R^{N}$ orthogonal to $\sigma_{1}+\cdots+\sigma_{N}$, the root system be the set of all vectors $\alpha\in \mathbf{E}$ for which $(\alpha,\alpha)=2$. It is obvious that the root system consists of all $\{\sigma_{i}-\sigma_{j}\}$, $i\neq j$ \cite{Hum72} for $A$-type gauge theory. Here we choose the representation
 \begin{equation}\label{}
    \mathcal{R}^{'}=V\otimes V^{*}\oplus V\otimes \mathcal{F}\oplus V\otimes \mathcal{F}^{'}
\end{equation}
which corresponds to the $H^{\text{max}}=U(L)\times U(L)\times U(1)$ global symmetry group. Here $V=\mathbf{C}^{N}$ is the $N$-dimensional fundamental representation, $\mathcal{F}\approx \mathbf{C}^{N_f}$, $\mathcal{F}^{'} \approx \mathbf{C}^{N_f}$ are the $N_f$-dimensional fundamental representations of the first and second $U(L)$ factors in the flavour group, and $\mathcal{L}$ is the standard one-dimensional representation of the global group $U(1)$. We have the effective potential of Robin boundary with a zero $\text{FI}$-term
\begin{equation}\label{13}
    \begin{aligned}
     W^{4d,\text{R}}_{\text{eff}}(\Phi)=&\sum_{j\ne  i}^N\dfrac{i\pi(\Phi_i-\Phi_j)^{3}}{3\tau \sigma}-\sum_{j\ne i}\dfrac{i\pi(1+\tau)(\Phi_i-\Phi_j)^{2}}{2\tau\sigma}\\  
     &-\sum_{j\ne  i}^N\dfrac{i\pi(\Phi_i-\Phi_j+m_{adj})^{3}}{3\tau \sigma}+\sum_{j\ne i}\dfrac{i\pi(1+\tau)(\Phi_i-\Phi_j+m_{adj})^{2}}{\tau\sigma}\\
 \end{aligned}
\end{equation}
\begin{equation*}
\begin{aligned}
     &-\sum_{j\ne i}^N\Big[\dfrac{i\pi (1+3\tau)(\Phi_i-\Phi_j+m_{adj})}{6\tau \sigma}-\dfrac{i\pi (1+3\tau)(\Phi_i-\Phi_j)}{6\tau \sigma}\Big]\\
     &+\dfrac{1}{2\pi i\sigma}\sum_{j\ne i}^N\sum_{n= 1}^{\infty}\left[\text{Li}_{2}(p^{n}e^{2\pi i (\Phi_i-\Phi_j+m_{adj})})-\text{Li}_{2}(p^{n}e^{2\pi i (\Phi_i-\Phi_j)})\right.\\
     &\qquad\left.-\text{Li}_{2}(p^{n-1}e^{-2\pi i(\Phi_i-\Phi_j+m_{adj})})+\text{Li}_{2}(p^{n-1}e^{-2\pi i(\Phi_i-\Phi_j)})\right]\\    
     &+\sum_{i=1}^N\sum_{a=1}^{N_f}\dfrac{i\pi  (\Phi_i+m_a)^{3}}{3\tau \sigma}-\sum_{i=1}^N\sum_{a=1}^{N_f}\dfrac{i\pi (1+\tau)(\Phi_i+m_a)^{2}}{2\tau \sigma}\\
     &+\sum_{i=1}^N\sum_{a=1}^{N_f}\Big[\dfrac{i\pi (1+3\tau)(\Phi_i+m_a)}{6 \tau \sigma}+\dfrac{i\pi (1+3\tau)(-\Phi_i+m_a^{'})}{6 \tau \sigma}\Big]\\
     &+\sum_{i=1}^N\sum_{a=1}^{N_f}\dfrac{i\pi (-\Phi_i+m^{'}_a)^{3}}{3\tau \sigma}-\sum_{i=1}^N\sum_{a=1}^{N_f}\dfrac{i\pi (1+\tau)(-\Phi_i+m_a^{'})^{2}}{2\tau \sigma}\\
     &+\frac{1}{2\pi i\sigma}\sum_{i=1}^N\sum_{a=1}^{N_f}\sum_{n=1}^{\infty}\left[\text{Li}_2(p^ne^{-2\pi i(-\Phi_i+m_a^{'})}-\text{Li}_2(p^{n-1}e^{2\pi i(-\Phi_i+m_a^{'})})\right.\\
     &\qquad\left. +\text{Li}_2(p^ne^{-2\pi i(\Phi_i+m_a)}-\text{Li}_2(p^{n-1}e^{2\pi i(\Phi_i+m_a)})\right]\\
\end{aligned}
\end{equation*}
Using the formula (\ref{12}) in 4D, we have the vacuum equations
\begin{equation}
\begin{aligned}
&e^{ph^A_{adj}}\prod_{j=1}^N\prod_{m=1}^{\infty}\frac{(1-p^mx_1)(1-p^{m-1}x_1^{-1})}{(1-p^mx_2)(1-p^{m-1}x_2^{-1})}\\
&=e^{ph^A_{fd}}\prod_{a=1}^{N_f}\prod_{m=1}^{\infty}\frac{(1-p^my_1)(1-p^{m+1}y_1^{-1})}{(1-p^my_2)(1-p^{m-1}y_2^{-1})}
\end{aligned}
\end{equation}
where $x_1=e^{2\pi i(\Phi_i-\Phi_j-m_{adj})}$, $x_2=e^{2\pi i(\Phi_i-\Phi_j+m_{adj})}$, $y_1=e^{2\pi i(\Phi_i-m_a)}$ and $y_2=e^{2\pi i(\Phi_i+m_a^{'})}$. The phases parameters are given by
\begin{equation*}
\begin{aligned}
&ph^A_{adj}=\sum_{j\neq i}^N\frac{-4\pi i(\Phi_i-\Phi_j)m_{adj}}{\tau}\\
&ph^A_{fd}=-\sum_{a=1}^{N_f}\left[\frac{2\pi i\Phi_i(m_a+m_a^{'})+i\pi(m_a^2-(m_a^{'})^2)-2\pi i(1+\tau)\Phi_i}{\tau}\right.\\
&\qquad \qquad ~~~ -\left.\frac{\pi i(1+\tau)(m_a-m_a^{'})}{\tau}\right]
\end{aligned}
\end{equation*}
The vacuum equations can be equivalently written as
\begin{equation}\label{77}
\begin{aligned}
&(-1)^{N_f-N-1}e^{ph^A_{adj}}\prod_{j=1}^N\frac{e^{\pi i(\Phi_i-\Phi_j+m_{adj})}}{e^{-\pi i(\Phi_i-\Phi_j-m_{adj})}}\frac{\sigma(\Phi_i-\Phi_j-m_{adj})}{\sigma(\Phi_i-\Phi_j+m_{adj})} \\
&=e^{ph^A_{fd}}\prod_{a=1}^{N_f}\frac{e^{\pi i (\Phi_i+m_a)}}{e^{-\pi i(\Phi_i-m_a^{'})}}\frac{\sigma (\Phi_i-m_a^{'})}{\sigma(\Phi_i+m_a)}
\end{aligned}
\end{equation}
Let 
\begin{equation*}
\begin{aligned}
&Ph_{A}=ph^A_{adj} - ph^A_{fd} \\
&=\sum_{j\neq i}^N\frac{-4\pi i(\Phi_i-\Phi_j)m_{adj}}{\tau}+\sum_{a=1}^{N_f}\left[\frac{2\pi i\Phi_i(m_a+m_a^{'})}{\tau}\right.\\
&\qquad \quad  + \left.\frac{i\pi(m_a^2-(m_a^{'})^2)-2\pi i(1+\tau)\Phi_i-\pi i(1+\tau)(m_a-m_a^{'})}{\tau}\right]\\
\end{aligned}
\end{equation*}
The derived formula $e^{Ph_{A}}$ exhibits precise agreement with the results in \cite{WZ23}, leading to the vacuum equation
\begin{equation}\label{14}
(-1)^{N_f-N-1}\prod_{j\neq i}^N\frac{\sigma(\Phi_i-\Phi_j-m_{\rm adj})}{\sigma(\Phi_i-\Phi_j+m_{\rm adj})}=\prod_{a=1}^{N_f}\frac{\sigma(\Phi_i-m'_a)}{\sigma(\Phi_i+m_a)}.
\end{equation}

By judiciously selecting mass parameters, one can construct the phase factor
\begin{equation}
Ph_{A} = 4\pi i l_1 u_j + 2i\phi
\end{equation}
the invariance requirement of (\ref{14}) is
\begin{equation}\label{}
\sum_{a=1}^{N_f}(m_a + m'_a) = -N m_{\rm adj},
\end{equation}
which exactly replicates the field content condition in 4D $\mathcal{N}=1$ gauge theories \cite{NS09a}.

The established correspondence dictionary
\begin{equation}\label{}
\begin{aligned}
\Phi_i &\leftrightarrow u_i, & m_{\rm adj} &= m_a \leftrightarrow \eta, \\
L &\leftrightarrow N_f, & N &\leftrightarrow \bar{M},
\end{aligned}
\end{equation}
provides complete identification between the BAEs (\ref{71}) and the gauge vacuum equations (\ref{77}).

\subsection{$B$-type gauge theory}

In the case of $SO(2N+1)$ gauge group, all the roots are given by $\{\pm \sigma_{i}\pm \sigma_{j}\}$ for all the possible combinations of $i<j$ and $\{\pm \sigma_{i}\}_{i=1}^{N}$. Therefore, the effective potential is given by
\begin{equation}
    \begin{aligned}
     W^{4d,\text{R}}_{\text{eff}}(\Phi)=&\sum_{i<j}\Big[\dfrac{i\pi(\pm \Phi_{i}\pm \Phi_{j})+m_{adj})^{3}}{3\tau \sigma}-\dfrac{i\pi(\pm \Phi_{i}\pm \Phi_{j})^{3}}{3\tau \sigma}\Big]\\
     &-\sum_{i<j}\Big[\dfrac{i\pi(1+\tau)(\pm \Phi_{i}\pm \Phi_{j})+m_{adj})^{2}}{2\tau\sigma}-\dfrac{i\pi(1+\tau)(\pm \Phi_{i}\pm \Phi_{j})^{2}}{2\tau\sigma}\Big]\\
     &+\sum_{i<j}\Big[\dfrac{i\pi (1+3\tau)(\pm \Phi_{i}\pm \Phi_{j}+m_{adj})}{6\tau \sigma}-\dfrac{i\pi (1+3\tau)(\pm \Phi_{i}\pm \Phi_{j})}{6\tau \sigma}\Big]\\
     &-\dfrac{1}{2\pi i\sigma}\sum_{i<j}\sum_{n= 1}^{\infty}\left[\text{Li}_{2}(p^{n}e^{2\pi i 
     (\pm \Phi_{i}\pm \Phi_{j})})-\text{Li}_{2}(p^{n-1}e^{-2\pi i(\pm \Phi_{i}\pm \Phi_{j})})\right.\\
     &\qquad \left. -\text{Li}_{2}(p^{n}e^{2\pi i (\pm \Phi_{i}\pm \Phi_{j}+m_{adj})})+\text{Li}_{2}(p^{n-1}e^{-2\pi i(\pm \Phi_{i}\pm \Phi_{j}+m_{adj})})\right]\\   
     &+\sum_{i=1}^N\sum_{a=1}^{N_f}\dfrac{i\pi (\pm \Phi_{i}+m_a)^{3}}{3\tau \sigma}-\sum_{i=1}^N\sum_{a=1}^{N_f}\dfrac{i\pi (1+\tau)(\pm \Phi_{i}+m_a)^{2}}{2\tau \sigma}\\
     &+\sum_{i=1}^N\sum_a^{N_f}\dfrac{i\pi(1+3\tau)(\pm \Phi_{i}+m_a)}{6 \tau \sigma}\\
     &+\frac{1}{2\pi i\sigma}\sum_{i=1}^N\sum_{a=1}^{N_f}\sum_{n=1}^{\infty}\left[\text{Li}_2(p^ne^{-2\pi i(\pm \Phi_{i}+m_a)})-\text{Li}_2(p^{n-1}e^{2\pi i(\pm \Phi_{i}+m_a)})\right]\\ 
     &+\sum_{i=1}^N\Big[\dfrac{i\pi(\pm 2\Phi_{i})^{3}}{3\tau \sigma}-\dfrac{i\pi(1+\tau)(\pm 2\Phi_{i})^{2}}{2\tau\sigma}+\dfrac{i\pi (1+3\tau)(\pm 2\Phi_{i})}{6\tau \sigma}\Big]\\ 
     &-\sum_{i=1}^N\dfrac{i\pi(2(\pm \Phi_{i}+m_{adj}))^{3}}{3\tau \sigma}+\sum_{i=1}^N\dfrac{i\pi(1+\tau)(2(\pm \Phi_{i}+m_{adj}))^{2}}{2\tau\sigma}\\
     &-\sum_{i=1}^N\dfrac{i\pi 2(1+3\tau)(\pm \Phi_{i}+m_{adj})}{6\tau \sigma}\\
     &+\dfrac{1}{2\pi i\sigma}\sum_{i=1}^N\sum_{n= 1}^{\infty}\left[\text{Li}_{2}(p^{n}e^{4\pi i 
     (\pm \Phi_{i}+m_{adj})})-\text{Li}_{2}(p^{n-1}e^{-4\pi i(\pm \Phi_{i}+m_{adj})})\right.\\
     &\qquad\left. -\text{Li}_{2}(p^{n}e^{4\pi i (\pm \Phi_{i})})+\text{Li}_{2}(p^{n-1}e^{-4\pi i(\pm \Phi_{i})})\right]\\  
\end{aligned}
\end{equation}

The vacuum equations are obtained by (\ref{12})
\begin{equation}\label{16}
\begin{aligned}
    &{\rm exp}\left(\frac{16\pi i \Phi_i m_{adj}}{\tau}+\sum_{a=1}^{N_f}\dfrac{i\pi(4\Phi_{i}m_a^f-2(1+\tau)\Phi_i)}{\tau}\right.\\
    &\left.~~~ ~~+\sum_{i<j}\frac{8i\pi \Phi_{i}m_{adj}-4i\pi(1+\tau) m_{adj}}{\tau})\right)\\
    &\times \prod_{i<j}\frac{(e^{2\pi i(-\Phi_{i}\pm \Phi_{j})};p)_{\infty}}{(e^{2\pi i(\Phi_{i}\pm \Phi_{j})};p)_{\infty}}\frac{(pe^{-2\pi i(-\Phi_{i}\pm \Phi_{j})};p)_{\infty}}{(pe^{2\pi i(\Phi_{i}\pm \Phi_{j})};p)_{\infty}}\\
    &\times \prod_{i<j}\frac{(e^{-2\pi i(\Phi_{i}\pm \Phi_{j}-m_{adj})};p)_{\infty}}{(e^{-2\pi i(\Phi_{i}\pm \Phi_{j}+m_{adj})};p)_{\infty}}\frac{(pe^{2\pi i(\Phi_{i}\pm \Phi_{j}-m_{adj})};p)_{\infty}}{(pe^{2\pi i(\Phi_{i}\pm \Phi_{j}+m_{adj})};p)_{\infty}}\\
    &\times \frac{(e^{-4\pi i\Phi_{i}};p)_{\infty}^2}{(pe^{-4\pi i\Phi_{i}};p)_{\infty}^2}\frac{(pe^{4\pi i\Phi_{i}};p)_{\infty}^2}{(e^{4\pi i\Phi_{i}};p)_{\infty}^2}\frac{(e^{4\pi i(\Phi_{i}-m_{adj})};p)_{\infty}^2}{(e^{4\pi i(\Phi_{i}+m_{adj})};p)_{\infty}^2}\frac{(p e^{-4\pi i(\Phi_{i}-m_{adj})};p)_{\infty}^2}{(pe^{-4\pi i(\Phi_{i}+m_{adj})};p)_{\infty}^2}\\
    &=\prod_{a=1}^{N_f}\frac{(e^{2\pi i(\Phi_{i}-m_{a})};p)_{\infty}}{(e^{2\pi i(\Phi_{i}+m_a)};p)_{\infty}}\frac{(p e^{-2\pi i(\Phi_{i}-m_a)};p)_{\infty}}{(p e^{-2\pi i(\Phi_{i}+m_a)};p)_{\infty}},\qquad i=1,\cdots,N
\end{aligned}
\end{equation}
The vacuum equations also can be 
\begin{equation}\label{}
\begin{aligned}
    &e^{ph^B_{adj}}\frac{e^{2\pi i(\Phi_{i}+m_{adj})}}{e^{-2\pi i(\Phi_{i}-m_{adj})}}\frac{\sigma(2(\Phi_{i}-m_{adj}))^2}{\sigma(2(\Phi_{i}+m_{adj}))^2}\prod_{i<j}^N\frac{e^{\pi i(\Phi_{i}\pm \Phi_{j}+m_{adj})}}{e^{-\pi i(\Phi_{i}\pm \Phi_{j}-m_{adj})}}\frac{\sigma(\Phi_{i}\pm \Phi_{j}-m_{adj})}{\sigma(\Phi_{i}\pm \Phi_{j}+m_{adj})}\\
    &=e^{ph^B_{fd}}\prod_{a=1}^{N_f}\frac{e^{-i\pi(\Phi_{i}+m_a)}}{e^{i\pi(\Phi_{i}-m_a)}}\frac{\sigma(\Phi_{i}-m_a)}{\sigma(\Phi_{i}+m_a)},\qquad i=1,\cdots,N
\end{aligned}
\end{equation}
The phases parameters are given by
$$ph^B_{adj}=\frac{16\pi i \Phi_i m_{adj}}{\tau}+\sum_{i<j}\frac{8i\pi \Phi_{i}m_{adj}-4i\pi(1+\tau) m_{adj}}{\tau}$$
$$ph^B_{fd}=-\sum_{a=1}^{N_f}\frac{i\pi(4\Phi_{i}m_a-2(1+\tau)\Phi_i)}{\tau}$$
Let us define the functional
\begin{equation}
\begin{aligned}
Ph_{B} &= ph^B_{adj} - ph^B_{fd} +\sum_{i<j}2\pi i(\Phi_i \pm \Phi_j)+4\pi i \Phi+2N_f\pi i \Phi_i \\
&=  \sum_{i<j} \biggl[ \frac{8i\pi \Phi_i m_{\rm adj} - 4i\pi(1+\tau)m_{\rm adj} + 2\pi i\tau(\Phi_i \pm \Phi_j)}{\tau} \biggr] \\
& ~~+ \sum_{a=1}^{N_f} \frac{i\pi(4\Phi_i m_a - 2\tau\Phi_i)}{\tau} + \frac{16\pi i\Phi_i m_{\rm adj} + 4\pi i\Phi_i\tau}{\tau}.
\end{aligned}
\end{equation}
Imposing the condition 
\begin{equation*}
Ph_{B} = 2\pi i k \quad (k \in \mathbb{Z})
\end{equation*}
we precisely derive the desired vacuum equations.
\begin{equation}\label{17}
\begin{aligned}
      &\frac{\sigma(2(\Phi_{i}-m_{adj}))^2}{\sigma(2(\Phi_{i}+m_{adj}))^2}\prod_{i<j}^N\frac{\sigma(\Phi_{i}\pm \Phi_{j}-m_{adj})}{\sigma(\Phi_{i}\pm \Phi_{j}+m_{adj})}=\prod_{a=1}^{N_f}\frac{\sigma(\Phi_{i}-m_a)}{\sigma(\Phi_{i}+m_a)},\qquad i=1,\cdots,N
\end{aligned}
\end{equation}

Actually, we can choose suitable mass parameters to let the factor $e^F$ be zero. Therefore we have the dictionary
\begin{equation}
\left\{
\begin{aligned}
&m_{adj}\leftrightarrow \eta\\
&m_a\leftrightarrow\frac{\eta}{2}, \qquad a=1,\cdots, 2L+1 \\
&m_a\leftrightarrow\eta, \qquad a=2L+2, 2L+3\\
&m_a\leftrightarrow-\eta+\frac{1}{2}, \qquad a=2L+4, 2L+5\\
&m_a\leftrightarrow-\eta+\frac{\tau}{2}, \qquad a=2L+6, 2L+7\\
&m_a\leftrightarrow-\eta-\frac{1+\tau}{2}, \qquad a=2L+8, 2L+9\\
&m_{a}\leftrightarrow\frac{\eta-1}{2},\qquad a=2L+10\\
&m_{a}\leftrightarrow\frac{\eta+\tau}{2},\qquad a=2L+11\\
&m_{a}\leftrightarrow\frac{\eta-1-\tau}{2},\qquad a=2L+12\\
\end{aligned}
\right.
\end{equation}
and
\begin{equation}
\begin{aligned}
&\Phi_i\leftrightarrow iu_i,\qquad N_f\leftrightarrow2L+12,\qquad N\leftrightarrow L
\end{aligned}
\end{equation} 
Through rigorous computation, we determine that the boundary parameters of the $\text{XYZ}$ spin chain should satisfy
\begin{equation}\label{59}
\begin{aligned}
    &\epsilon_1^{+}=\epsilon_2^{+}=\epsilon_3^{-}=1,\qquad \epsilon_1^{-}=\epsilon_2^{-}=\epsilon_3^{+}=-1\\
    &\alpha_1^{+}=\alpha_1^{-}=\frac{1}{2},\qquad \alpha_2^{+}=\alpha_2^{-}=\frac{\tau}{2}, \qquad \alpha_3^{+}=\alpha_3^{-}=\frac{1+\tau}{2},
\end{aligned}
\end{equation}
We thus establish the pivotal result: the vacuum equation (\ref{17}) satisfies the BAE (\ref{8}).

\subsection{$C$-type gauge theory}

For $\text{Sp}(2N)$ gauge theory, all the roots are given by $\{\pm \sigma_{i}\pm \sigma_{j}\}$ for $i<j$ and $\{\pm 2\sigma_{i}\}_{i=1}^{N}$. Hence we get the effective superpotential by using the formula (\ref{12})
\begin{equation}
    \begin{aligned}
     W^{4d,\text{R}}_{\text{eff}}(\Phi)=&-\sum_{i<j}\dfrac{i\pi(\pm \Phi_{i}\pm \Phi_{j})^{3}}{3\tau \sigma}+\sum_{i<j}\dfrac{i\pi(1+\tau)(\pm \Phi_{i}\pm \Phi_{j})^{2}}{2\tau\sigma}\\
       \end{aligned}
\end{equation}
\begin{equation*}
\begin{aligned}
     &-\sum_{i<j}\dfrac{i\pi (1+3\tau)(\pm \Phi_{i}\pm \Phi_{j})}{6\tau \sigma}\\
     &+\frac{1}{2\pi i\sigma}\sum_{i<j}\sum_{k=1}^{\infty}\left[\frac{e^{2\pi ik(\pm \Phi_{i}\pm \Phi_{j})}}{k^2(1-p^k)}-\frac{p^ke^{-2\pi ik(\pm \Phi_{i}\pm \Phi_{j})}}{k^2(1-p^k)}\right]\\
     &+\sum_{i<j}\dfrac{i\pi(\pm \Phi_{i}\pm \Phi_{j})+m_{adj})^{3}}{3\tau \sigma}-\sum_{i<j}\dfrac{i\pi(1+\tau)(\pm \Phi_{i}\pm \Phi_{j})+m_{adj})^{2}}{2\tau\sigma}\\
     &+\sum_{i<j}\dfrac{i\pi (1+3\tau)(\pm \Phi_{i}\pm \Phi_{j})+m_{adj})}{6\tau \sigma}\\
     &-\frac{1}{2\pi i\sigma}\sum_{i<j}\sum_{k=1}^{\infty}\left[\frac{e^{2\pi ik(\pm \Phi_{i}\pm \Phi_{j}+m_{adj})}}{k^2(1-p^k)}-\frac{p^ke^{-2\pi ik(\pm \Phi_{i}\pm \Phi_{j}+m_{adj})}}{k^2(1-p^k)}\right]\\
     &+\sum_{i=1}^N\sum_{a=1}^{N_f}\dfrac{i\pi (\pm \Phi_{i}+m_a)^{3}}{3\tau \sigma}-\sum_{i=1}^N\sum_{a=1}^{N_f}\dfrac{i\pi (1+\tau)(\pm \Phi_{i}+m_a)^{2}}{2\tau \sigma}\\
     &+\sum_{i=1}^N\sum_{a=1}^{N_f}\dfrac{i\pi(1+3\tau)(\pm \Phi_{i}+m_a)}{6 \tau \sigma}\\
     &+\frac{1}{2\pi i\sigma}\sum_{i=1}^N\sum_{a=1}^{N_f}\sum_{k=1}^{\infty}\left[\frac{p^ke^{-2\pi ik(\pm \Phi_i+m_a)}}{k^2(1-p^k)}-\frac{e^{2\pi ik(\pm \Phi_i+m_a)}}{k^2(1-p^k)}\right]\\
     &-\sum_{i=1}^N\dfrac{i\pi(\pm \Phi_{i})^{3}}{3\tau \sigma}+\sum_{i=1}^N\dfrac{i\pi(1+\tau)(\pm \Phi_{i})^{2}}{2\tau\sigma}-\sum_{i=1}^N\dfrac{i\pi (1+3\tau)(\pm \Phi_{i})}{6\tau \sigma}\\
     &+\dfrac{1}{2\pi i\sigma}\sum_{i=1}^N\sum_{k= 1}^{\infty}\left[\text{Li}_{2}(p^{n}e^{2\pi i 
     (\pm \Phi_{i}})-\text{Li}_{2}(p^{n-1}e^{-2\pi i(\pm \Phi_{i})})\right]\\
     &-\sum_{i=1}^N\dfrac{i\pi(\pm \Phi_{i}+m_{adj})^{3}}{3\tau \sigma}+\sum_{i=1}^N\dfrac{i\pi(1+\tau)(\pm \Phi_{i}+m_{adj})^{2}}{2\tau\sigma}\\
     &-\sum_{i=1}^N\dfrac{i\pi (1+3\tau)(\pm \Phi_{i}+m_{adj})}{6\tau \sigma}\\
     &+\dfrac{1}{2\pi i\sigma}\sum_{i=1}^N\sum_{k= 1}^{\infty}\left[\frac{e^{2\pi ik(\pm \Phi_i+m_{adj})}}{k^2(1-p^k)}-\frac{p^ke^{-2\pi ik(\pm \Phi_i+m_{adj})}}{k^2(1-p^k)}\right]\\
\end{aligned}
\end{equation*}

The vacuum equations are 
\begin{equation}\label{}
\begin{aligned}
&e^{ph^C_{adj}}\frac{e^{\pi i(\Phi_{i}+m_{adj})}}{e^{-\pi i(\Phi_{i}-m_{adj})}}\frac{\sigma(\Phi_{i}-m_{adj})}{\sigma(\Phi_{i}+m_{adj})}\prod_{i<j}\frac{e^{\pi i(\Phi_{i}\pm \Phi_{j}+m_{adj})}}{e^{-\pi i(\Phi_{i}\pm \Phi_{j}-m_{adj})}}\frac{\sigma(\Phi_{i}\pm \Phi_{j}-m_{adj})}{\sigma(\Phi_{i}\pm \Phi_{j}+m_{adj})}\\
    =&e^{ph^C_{fd}}\prod_{a=1}^{N_f}\frac{e^{-i\pi (\Phi_{i}+m_a)}}{e^{i\pi (\Phi_{i}-m_a)}}\frac{\sigma(\Phi_{i}-m_a)}{\sigma(\Phi_{i}+m_a)},\qquad i=1,\cdots,N
\end{aligned}
\end{equation}
The phases parameters are given by
$$ph^C_{adj}=\frac{4\pi i \Phi_i m_{adj}}{\tau}+\sum_{i<j}\frac{8i\pi \Phi_{i}m_{adj}-4i\pi(1+\tau) m_{adj}}{\tau}$$
$$ph^C_{fd}=-\sum_{a=1}^{N_f}\frac{i\pi(4\Phi_{i}m_a-2(1+\tau)\Phi_i)}{\tau}$$
Define the functional 
\begin{equation}
\begin{aligned}
Ph_{C}&=ph^C_{adj} - ph^C_{fd} +\sum_{i<j}2\pi i(\Phi_i \pm \Phi_j)+2\pi i \Phi+2N_f\pi i \Phi_i  \\
&= \sum_{i<j} \biggl[ \frac{8\pi i \Phi_i m_{\rm adj} - 4\pi i(1+\tau)m_{\rm adj} + 2\pi i\tau(\Phi_i \pm \Phi_j)}{\tau} \biggr] \\
& ~~+ \sum_{a=1}^{N_f} \frac{\pi i(4\Phi_i m_a - 2(1+\tau)\Phi_i + 2\Phi_i\tau)}{\tau} + \frac{4\pi i\Phi_i m_{\rm adj} + 2N_f\pi i\tau\Phi_i}{\tau}
\end{aligned}
\end{equation}
Imposing the condition  
\begin{equation*}
Ph_{C}= 2\pi i k \quad (k \in \mathbb{Z}),
\end{equation*}
we rigorously derive the vacuum equations:
\begin{equation}\label{19}
\begin{aligned}
\frac{\sigma(\Phi_{i}-m_{\rm adj})}{\sigma(\Phi_{i}+m_{\rm adj})} \prod_{i<j} \frac{\sigma(\Phi_i \pm \Phi_j - m_{\rm adj})}{\sigma(\Phi_i \pm \Phi_j + m_{\rm adj})} = \prod_{a=1}^{N_f} \frac{\sigma(\Phi_i - m_a)}{\sigma(\Phi_i + m_a)}, \quad i=1,\dots,N
\end{aligned}
\end{equation}

Comparing the BAE (\ref{8}) and the vacuum equations (\ref{19}), we have the dictionary
\begin{equation}
\left\{
\begin{aligned}
&m_{adj}\leftrightarrow \eta\\
&m_a\leftrightarrow \frac{\eta}{2}, \qquad a=1,\cdots, 2L+1 \\
&m_a\leftrightarrow -\eta, \qquad a=2L+2 \\
&m_{a}\leftrightarrow \frac{\eta-1}{2},\qquad a=2L+3\\
&m_{a}\leftrightarrow \frac{\eta+\tau}{2},\qquad a=2L+4\\
&m_{a}\leftrightarrow \frac{\eta-1-\tau}{2},\qquad a=2L+5\\
\end{aligned}
\right.
\end{equation}
and
\begin{equation}
\Phi_i\leftrightarrow iu_i,\qquad N_f\leftrightarrow2L+5,\qquad N\leftrightarrow L
\end{equation}
The boundary parameters are the same to (\ref{59}). Then we get the Bethe/gauge correspondence for $C$-type gauge theory: $(\ref{19})\Longleftrightarrow (\ref{8})$.

\subsection{$D$-type gauge theory}

In the case of $\text{SO}(2N)$ gauge group, all the roots are given by $\{\pm \sigma_{i}\pm \sigma_{j}\}$ for all $1\le i< j\le N$. So the effective superpotential can be written as
\begin{equation}
    \begin{aligned}
     W^{4d,\text{R}}_{\text{eff}}(\Phi)=&-\sum_{i<j}\dfrac{i\pi(\pm \Phi_{i}\pm \Phi_{j})^{3}}{3\tau \sigma}+\sum_{i<j}\dfrac{i\pi(1+\tau)(\pm \Phi_{i}\pm \Phi_{j})^{2}}{2\tau\sigma}\\
     &-\sum_{i<j}\dfrac{i\pi (1+3\tau)(\pm \Phi_{i}\pm \Phi_{j})}{6\tau \sigma}\\
     &+\frac{1}{2\pi i\sigma}\sum_{i<j}\sum_{k=1}^{\infty}\left[\frac{e^{2\pi ik(\pm \Phi_{i}\pm \Phi_{j})}}{k^2(1-p^k)}-\frac{p^ke^{-2\pi ik(\pm \Phi_{i}\pm \Phi_{j})}}{k^2(1-p^k)}\right]\\
     &+\sum_{i<j}\dfrac{i\pi(\pm \Phi_{i}\pm \Phi_{j})+m_{adj})^{3}}{3\tau \sigma}-\sum_{i<j}\dfrac{i\pi(1+\tau)(\pm \Phi_{i}\pm \Phi_{j})+m_{adj}))^{2}}{2\tau\sigma}\\
     &+\sum_{i<j}\dfrac{i\pi (1+3\tau)(\pm \Phi_{i}\pm \Phi_{j})+m_{adj}))}{6\tau \sigma}\\
     &-\frac{1}{2\pi i\sigma}\sum_{i<j}\sum_{k=1}^{\infty}\left[\frac{e^{2\pi ik(\pm \Phi_{i}\pm \Phi_{j}+m_{adj})}}{k^2(1-p^k)}-\frac{p^ke^{-2\pi ik(\pm \Phi_{i}\pm \Phi_{j}+m_{adj})}}{k^2(1-p^k)}\right]\\
     &+\sum_{i=1}^N\sum_{a=1}^{N_f}\dfrac{i\pi (\pm \Phi_{i}i+m_a)^{3}}{3\tau \sigma}-\sum_{i=1}^N\sum_{a=1}^{N_f}\dfrac{i\pi (1+\tau)(\pm \Phi_{i}+m_a)^{2}}{2\tau \sigma}\\
     &+\sum_{i=1}^N\sum_{a=1}^{N_f}\dfrac{i\pi(1+3\tau)(\pm \Phi_{i}+m_a)}{6 \tau \sigma}\\
     &+\frac{1}{2\pi i\sigma}\sum_{i=1}^N\sum_{a=1}^{N_f}\sum_{k=1}^{\infty}\left[\frac{p^ke^{-2\pi ik\mu_a(\pm \Phi_i+m_a)}}{k^2(1-p^k)}-\frac{e^{2\pi ik(\pm \Phi_i+m_a)}}{k^2(1-p^k)}\right]\\
\end{aligned}
\end{equation}
One get the vacuum equations by using (\ref{12})
\begin{equation}\label{}
\begin{aligned}
     &e^{ph^D_{adj}}\prod_{i<j}\frac{e^{\pi i(\Phi_{i}\pm \Phi_{j}+m_{adj})}}{e^{-\pi i(\Phi_{i}\pm \Phi_{j}-m_{adj})}}\frac{\sigma(\Phi_{i}\pm \Phi_{j}-m_{adj})}{\sigma(\Phi_{i}\pm \Phi_{j}+m_{adj})}\\
    =&e^{ph^D_{bf}}\prod_{a=1}^{N_f}\frac{e^{-i\pi (\Phi_{i}+m_a)}}{e^{i\pi (\Phi_{i}-m_a)}}\frac{\sigma(\Phi_{i}-m_a)}{\sigma(\Phi_{i}+m_a)},\qquad i=1,\cdots,N
\end{aligned}
\end{equation}
The phases parameters are given by
$$ph^D_{adj}=\sum_{i<j}\frac{8i\pi \Phi_{i}m_{adj}-4i\pi(1+\tau) m_{adj}}{\tau}$$
$$ph^D_{fd}=-\sum_{a=1}^{N_f}\frac{i\pi(4\Phi_{i}m_a-2(1+\tau)\Phi_i)}{\tau}$$
Define
\begin{equation*}
\begin{aligned}
Ph_{D}&=ph^D_{adj} - ph^D_{fd}+\sum_{i<j}2\pi i(\Phi_i \pm \Phi_j)+2N_f\pi i \Phi_i\\
&=\sum_{i<j}\frac{8i\pi \Phi_{i}m_{adj}-4i\pi(1+\tau) m_{adj}+2i\pi \tau(\Phi_i\pm \Phi_j)}{\tau}\\
&~~+\sum_{a=1}^{N_f}\dfrac{i\pi(4\Phi_{i}m_a-2(1+\tau)\Phi_i+2\Phi_i\tau)}{\tau}
\end{aligned}
\end{equation*}
Imposing the condition  
\begin{equation*}
Ph_{D} = 2\pi i k \quad (k \in \mathbb{Z}),
\end{equation*}
we derive the following vacuum equations
\begin{equation}\label{20}
\begin{aligned}
     & \frac{\sigma(\Phi_{i}\pm \Phi_{j}-m_{adj})}{\sigma(\Phi_{i}\pm \Phi_{j}+m_{adj})}=\prod_{a=1}^{N_f}\frac{\sigma(\Phi_{i}-m_a)}{\sigma(\Phi_{i}+m_a)},\qquad i=1,\cdots,N
\end{aligned}
\end{equation}
Hence we set the dictionary
\begin{equation}
\left\{
\begin{aligned}
&m_{adj}\leftrightarrow \eta\\
&m_a^f\leftrightarrow \frac{\eta}{2}, \qquad a=1,\cdots, 2L+1 \\
&m_{a}^f\leftrightarrow \frac{\eta-1}{2},\qquad a=2L+2\\
&m_{a}^f\leftrightarrow \frac{\eta+\tau}{2},\qquad a=2L+3\\
&m_{a}^f\leftrightarrow\frac{\eta-1-\tau}{2},\qquad a=2L+4\\
\end{aligned}
\right.
\end{equation}
and 
\begin{equation}
\Phi_i\leftrightarrow iu_i,\qquad N_f\leftrightarrow2L+4,\qquad N\leftrightarrow L
\end{equation}
Here we can choose suitable mass parameters to let the factor $e^{Ph_{D}}$ be zero. Based on the above, we get the vacuum equations (\ref{20}) correspond to the BAE (\ref{8}) with the boundary parameters (\ref{59}).

\section{Conclusion and discussion}\label{s5}

This paper establishes a systematic correspondence between 4D $\mathcal{N}=1$ supersymmetric gauge theories and the $\text{XYZ}$ quantum spin chain with the most generic boundary conditions. We explicitly construct parameter dictionaries governing this duality across distinct gauge group types (A/B/C/D), revealing gauge-group-dependent boundary condition selection rules. For $A$-type supersymmetry gauge theory, the vacuum equations are shown to correspond to the BAEs of closed boundary condition. Conversely, we show the correspondence between the vacuum equation of the $\rm BCD$-type gauge theory and the BAE with open boundary - a mathematically inevitable consequence arising from the gauge group representations. Our approach leverages a unified methodology to compute effective superpotentials across dimensional hierarchies of supersymmetric gauge theories (2D/3D/4D), thereby establishing an explicit correspondence between these gauge-theoretic frameworks and quantum integrable spin chain systems. Through systematic analysis of vacuum solution manifolds and BAEs, we rigorously identify duality-satisfying parameter regimes, revealing a profound interplay between gauge dynamics and quantum integrability."The approach developed here to derive the duality between the $\rm XYZ$ spin chain and the 4D $\mathcal{N}=1$ supersymmetric $\rm{ABCD}$ gauge theory ‌works‌ for cases involving exceptional gauge groups. For brevity, we omit the cumbersome expressions in this space.

The effective superpotentials come from the partition function in supersymmetric gauge theories. If we consider the root structure of the corresponding Lie algebra of the gauge group, it implies some relation between geometry and representation theory, such as \cite{OR11}. On the other hand, quiver gauge theory is also a direction worth studying, and our method can be applied in parallel to quiver gauge theory.

\backmatter


\bmhead{Acknowledgements}

The financial support from the Natural Science Foundation of China (NSFC, Grants 11775299) is gratefully acknowledged from one of the authors (Ding). The second author thanks for Professor Shuai Guo for providing a supportive and enjoyable working environment.

\section*{Declarations}

\begin{itemize}
\item The authors declare that they have no known competing financial interests.
\item This study is a theoretical analysis and does not generate new experimental data.  
\end{itemize}

\begin{appendices}

\section{3D $\rm XXZ$ spin chain with general open boundary}\label{secA1}

In this Appendix, we aim to exhibit the Bethe/Gauge correspondence between 3D $\mathcal{N}=2$ gauge theory on $D^2\times S^1$ and the $\rm XXZ$ spin chain with off-diagonal boundary condition.

We will first recall the effective superpotential of 3d $\mathcal{N}=2$ gauge theory, as we gave in previous papers \cite{DZ23a,DZ23c}. In \cite{DZ23c}, we chose the representation (\ref{10}) which corresponds to the theory with the $H^{\text{max}}=U(N_f)\times U(N_f)$ global symmetry group. Similarly, $V=\mathbf{C}^{N}$ is the $N$-dimensional fundamental representation, $\mathcal{F}\approx \mathbf{C}^{L}$ is the $L$-dimensional fundamental representations. The effective superpotential for this theory was given by
\begin{equation}\label{}
\begin{aligned}
   W^{3d}_{\text{eff}}(\sigma,m)=&-\dfrac{1}{\beta^{2}}\sum_{\alpha \in \Delta}\text{Li}_{2}(e^{\frac{4}{\alpha_{i}^{2}}i\alpha\cdot \sigma})+\dfrac{1}{4\beta_{2}}\sum_{\alpha\in \Delta}(\frac{4}{\alpha_{i}^{2}}\alpha\cdot \sigma)^{2}\\
&+\dfrac{1}{\beta_{2}}\sum_{w \in \mathcal{R}}\sum_{a=1}^{N_{f}}\text{Li}_{2}(e^{2(-iw\cdot \sigma-im_{a}-i\beta_{2}\tilde{c})})\\
&-\dfrac{1}{4\beta_{2}}\sum_{w\in \mathcal{R}}\sum_{a=1}^{N_{f}}[2(w\cdot \sigma+m_{a}+\beta_{2}\tilde{c})]^{2}\\
    \end{aligned}
\end{equation}
where $\beta_{2}$ is the $U(1)$ charge fugacity.

For the $B$-type gauge theory, we have the vacuum equation
\begin{equation}\label{101}
 \dfrac{\text{sin}^{2}(\sigma_{j}-m_{\text{adj}})\text{cos}^{2}(\sigma_{j}-m_{\text{adj}})}{\text{sin}^{2}(\sigma_{j}+m_{\text{adj}})\text{cos}^{2}(\sigma_{j}+m_{\text{adj}})}\prod_{j\neq k}\dfrac{\text{sin}(\sigma_{j}\pm \sigma_{k}-m_{\text{adj}})}{\text{sin}(-\sigma_{j}\pm \sigma_{k}-m_{\text{adj}})}\prod_{a=1}^{N_{f}}\dfrac{\text{sin}(\sigma_{j}-m_{a})}{\text{sin}(-\sigma_{j}-m_{a})}=1
\end{equation}
The boundary parameters of the $\text{XXZ}$ spin chain should satisfy
\begin{equation}\label{}
\begin{aligned}
    &\epsilon_1^{\pm}=\epsilon_2^{\pm}=\epsilon_3^{\pm}=1\\
    &\alpha_1^{+}=0,\quad  \alpha_1^{-}=\frac{1}{2},\quad \alpha_2^{\pm}=\frac{\eta}{2},\quad \alpha_3^{\pm}=\frac{\eta+1}{2}
\end{aligned}
\end{equation}
The dictionary 
\begin{equation}\label{102}
\begin{aligned}
&\pi ui \longleftrightarrow \sigma ,\quad \pi \eta \longleftrightarrow m_{\text{adj}}\\
& L \longleftrightarrow N,\quad 2L\longleftrightarrow N_{f} \\
&-\dfrac{\pi \eta}{2} \longleftrightarrow  m_{a}
 \end{aligned}
\end{equation}
satisfies the duality of the BAE (\ref{100}) and the vacuum equation (\ref{101}).

For the $C$-type gauge theory, we have the vacuum equation
\begin{equation}\label{103}
\dfrac{\text{sin}(\sigma_{j}-\frac{m_{\text{adj}}}{2})}{\text{sin}(\sigma_{j}+\frac{m_{\text{adj}}}{2})}\prod_{j\neq k}^{N}\dfrac{\text{sin}(\sigma_{j}\pm \sigma_{k}-m_{\text{adj}})}{\text{sin}(-\sigma_{j}\pm \sigma_{k}-m_{\text{adj}})}\prod_{a=1}^{N_{f}}\dfrac{\text{sin}(\sigma_{j}-m_{a})}{\text{sin}(-\sigma_{j}-m_{a})}=1
\end{equation}
The boundary parameters of the $\text{XXZ}$ spin chain should satisfy
\begin{equation}\label{}
\begin{aligned}
    &\epsilon_1^{\pm}=\epsilon_2^{\pm}=\epsilon_3^{\pm}=1\\
    &\alpha_1^{+}=\alpha_2^{+}=0,\quad  \alpha_1^{-}=\frac{1}{2},\quad \alpha_2^{-}=\alpha_3^{\pm}=-\frac{\eta}{2}
\end{aligned}
\end{equation}
The dictionary (\ref{102}) satisfies the duality of the BAE (\ref{100}) and the vacuum equation (\ref{103}).

For the $D$-type gauge theory, we have the vacuum equation
\begin{equation}\label{104}
\prod_{j\neq k}^{N}\dfrac{\text{sin}(\sigma_{j}\pm \sigma_{k}-m_{\text{adj}})}{\text{sin}(-\sigma_{j}\pm \sigma_{k}-m_{\text{adj}})}\prod_{a=1}^{N_{f}}\dfrac{\text{sin}(\sigma_{j}-m_{a})}{\text{sin}(-\sigma_{j}-m_{a})}=1
\end{equation}
The boundary parameters of the $\text{XXZ}$ spin chain should satisfy
\begin{equation}\label{}
\begin{aligned}
    &\epsilon_1^{\pm}=\epsilon_2^{\pm}=\epsilon_3^{\pm}=1\\
    &\alpha_1^{+}=0,\quad  \alpha_1^{-}=\frac{1}{2},\quad \alpha_2^{\pm}=\alpha_3^{\pm}=-\frac{\eta}{2}
    \end{aligned}
\end{equation}
The dictionary (\ref{102}) satisfies the duality of the BAE (\ref{100}) and the vacuum equation (\ref{104}).

The Bethe/Gauge correspondence between known 3D vacuum equations (\cite{KZ21,WZ23}) — including those in \cite{DZ23a} — and the BAEs (\ref{100}) ‌can be established analogously to the procedure described above.




\end{appendices}


\end{document}